\documentclass[sn-mathphys,Numbered]{sn-jnl}

\usepackage{graphicx}%
\usepackage{multirow}%
\usepackage{amsmath,amssymb,amsfonts}%
\usepackage{amsthm}%
\usepackage{mathrsfs}%
\usepackage[title]{appendix}%
\usepackage{xcolor}%
\usepackage{textcomp}%
\usepackage{manyfoot}%
\usepackage{booktabs}%
\usepackage{algorithm}%
\usepackage{algorithmicx}%
\usepackage{algpseudocode}%
\usepackage{listings}%
\usepackage{tcolorbox}%
\usepackage{makecell}
\usepackage{hyperref}

\theoremstyle{thmstyleone}%
\theoremstyle{thmstyletwo}%

\theoremstyle{thmstylethree}%

\begin{document}

\title[Article Title]{DNAGPT: A Generalized Pre-trained Tool for Versatile DNA Sequence Analysis Tasks}


\author[1,2,4]{\fnm{Daoan} \sur{Zhang}}\email{daoan.zhang@rochester.edu}
\author[2,3]{\fnm{Weitong} \sur{Zhang}}\email{weitzhang6-c@my.cityu.edu.hk}
\author[2]{\fnm{Yu} \sur{Zhao}}\email{louisyuzhao@tencent.com}
\author*[1]{\fnm{Jianguo} \sur{Zhang}}\email{zhangjg@sustech.edu.cn}
\author*[2]{\fnm{Bing} \sur{He}}\email{owenbhe@tencent.com}
\author*[2]{\fnm{Chenchen} \sur{Qin}}\email{chenchenqin@tencent.com}
\author*[2]{\fnm{Jianhua} \sur{Yao}}\email{jianhuayao@tencent.com}

\affil[1]{\orgname{Southern University of Science and Technology}}
\affil[2]{\orgname{Tencent AI Lab, Shenzhen, China}}
\affil[3]{\orgname{City University of Hong Kong}}
\affil[4]{\orgname{University of Rochester}}





\abstract{

Pre-trained large language models demonstrate potential in extracting information from DNA sequences, yet adapting to a variety of tasks and data modalities remains a challenge. To address this, we propose DNAGPT, a generalized DNA pre-training model trained on over 200 billion base pairs from all mammals. By enhancing the classic GPT model with a binary classification task (DNA sequence order), a numerical regression task (guanine-cytosine content prediction), and a comprehensive token language, DNAGPT can handle versatile DNA analysis tasks while processing both sequence and numerical data. Our evaluation of genomic signal and region recognition, mRNA abundance regression, and artificial genomes generation tasks demonstrates DNAGPT's superior performance compared to existing models designed for specific downstream tasks, benefiting from pre-training using the newly designed model structure.
}




\maketitle

\section{Introduction}



DNA serves as the essential blueprint of life, encompassing the comprehensive instruction manual that guides an organism through growth, development, survival, and reproduction. The Human Genome Project has advanced our understanding of life by decoding the DNA code, leading to ongoing research in DNA interpretation, biological processes, disease detection, and the redesign of life, which can be applied in bacterial and mammalian cell engineering for both diagnostics and therapeutics by synthetic biological technologies such as the CRISPR-Cas system \cite{tipper1965mechanism, giege2012structure, chen2023single, mcnerney2021theranostic}. As the most fundamental information in biology, DNA sequences contain rich biological information\cite{shendure2008next}, especially those with large non-coding regions \cite{andolfatto2005adaptive} that remain unexplored and are particularly worth investigating. The considerable diversity, vast volume, and intricate relationships within biological information pose challenges in the analysis and comprehension of such data. For example, as the basic functional unit of DNA sequence, a single gene, among the estimated 100,000 genes present in the human genome,\cite{gibbs2020human} can be characterized from different aspects: it can be represented by nucleotide sequences \cite{brosius1978complete}, its expression level in different cells may vary greatly due to the influence of factors such as its non-coding region, cell type, or environment \cite{morley2004genetic}, moreover, it can be translated into proteins with different abundance levels under different circumstances \cite{pearson2006gene}. Consequently, DNA sequence research requires the integration of sequencing data, represented by DNA sequences, and expression data, represented by numerical values of abundance.

Recently, the advent of foundation models \cite{luo2022biogpt, kirillov2023segment, touvron2023llama} has revolutionized natural language understanding \cite{bommasani2021opportunities} through the pre-training of generalized models on large-scale datasets, which can be fine-tuned for various downstream tasks. Inspired by this, pre-trained models have been employed to uncover the hidden information within DNA sequences \cite{ji2021dnabert,dalla2023nucleotide}. However, as mentioned above, DNA analysis tasks have various forms that involve both sequence and numerical data as input and output \cite{yelmen2021creating, andrews2023mammalian, wang2023deepbio} which are difficult to tackle in one language-based model \cite{encode2012integrated, chen2022capturing, wang2022towards, lee2022learning}. The previous attempts, DNABERT \cite{ji2021dnabert} as well as Nucleotide Transformers (NT) \cite{dalla2023nucleotide}, involved pre-training on the genome data followed by fine-tuning on the downstream datasets based on task-specific heads, separately handling attribute prediction tasks like the recognition of genomic signals and regions (GSR) tasks \cite{kalkatawi2019deepgsr, guo2022context, zhu2023gsrnet} and generation tasks like reconstructing human genetic variants \cite{dalla2023nucleotide}. In addition, during pre-training, the previously mentioned pre-trained models only used DNA sequences and did not consider numerical data, making it unsuitable for tasks that involve numerical input or output such as the regression of mRNA abundance from the DNA sequence \cite{agarwal2020predicting}. These weaknesses severely limit the generalization of various tasks and fail to propose a generalized model that seamlessly integrates DNA sequence-relevant tasks. Also, unifying those intricate and diverse data types and task paradigms can reduce unnecessary algorithm design effort while allowing more tasks to benefit from pre-training, further paving the way for more profound discoveries and insights in DNA sequence analysis. Therefore, a generalized pre-training model is needed to fully extract and utilize DNA information, which adapts to various DNA-related downstream tasks, to gain a comprehensive perspective on DNA, accelerate research and production processes, improve research accuracy, and avoid the waste of resources caused by repeated research.

Constructing such a generalized pre-trained model for DNA sequences requires consideration from two aspects: (1) How to coherently process different data types (sequence and number) in both the pre-training and testing stages? (2) How to establish a common pipeline for different tasks?  
In this study, we introduce DNAGPT, a generalized pre-trained model for DNA analysis, where a  multi-task pre-training strategy and a  novel token  language  are proposed to answer the above two questions. In addition to the auto-regression pre-training task in the classic GPT model, we add a binary classification pre-training task (DNA sequence order) and a numerical regression pre-training task (guanine-cytosine content prediction) in the pre-training stage to help the model to better understand DNA sequence data and numerical data. For the DNA sequence order prediction, we randomly flip the input DNA sequence and let the model predict whether the flip operation has been performed or not. For the guanine-cytosine (GC) content prediction, we randomly extract a segment of the sequence from the input and then have the model calculate and output the GC content value for this segment. We modify the GPT architecture with corresponding embedding layers and encoding heads for both sequence and numerical input and outputs so that they can be processed and trained in the same framework. We also design a comprehensive token language to encode sequence, number, and task-related information in the same token space.
Furthermore, in order to better learn the sequence conservation and diversity across species, we utilize reference genomes  \cite{cunningham2022ensembl} from all the mammals for pre-training, with a total data size exceeding 200 billion base pairs (bps). 

After pre-training, we tested and evaluated the functionalities, capabilities and performance of the DNAGPT on a diverse panel of prediction, regression, and generation tasks. We began from GSR prediction task \cite{kalkatawi2019deepgsr} to assess the sensitivity of the model to specific sites. The results demonstrated that the DNAGPT can not only compete with state-of-the-art methods but also accurately identify pivotal regions within the input sequence. After that, DNAGPT achieved better results compared with conventional methods on mRNA abundance assessment task \cite{agarwal2020predicting} with a mixture input of tensors and DNA sequences and output the corresponding mRNA abundance values. We further examined whether DNAGPT can produce pseudo DNA sequences \cite{yelmen2021creating}, the results from various metrics proved that the DNAGPT surpassed traditional GAN and RBM models in terms of maintaining certain biological properties and features discovered in natural genomic sequences.

\begin{figure}[!t]%
\centering
\includegraphics[width=0.9\textwidth]{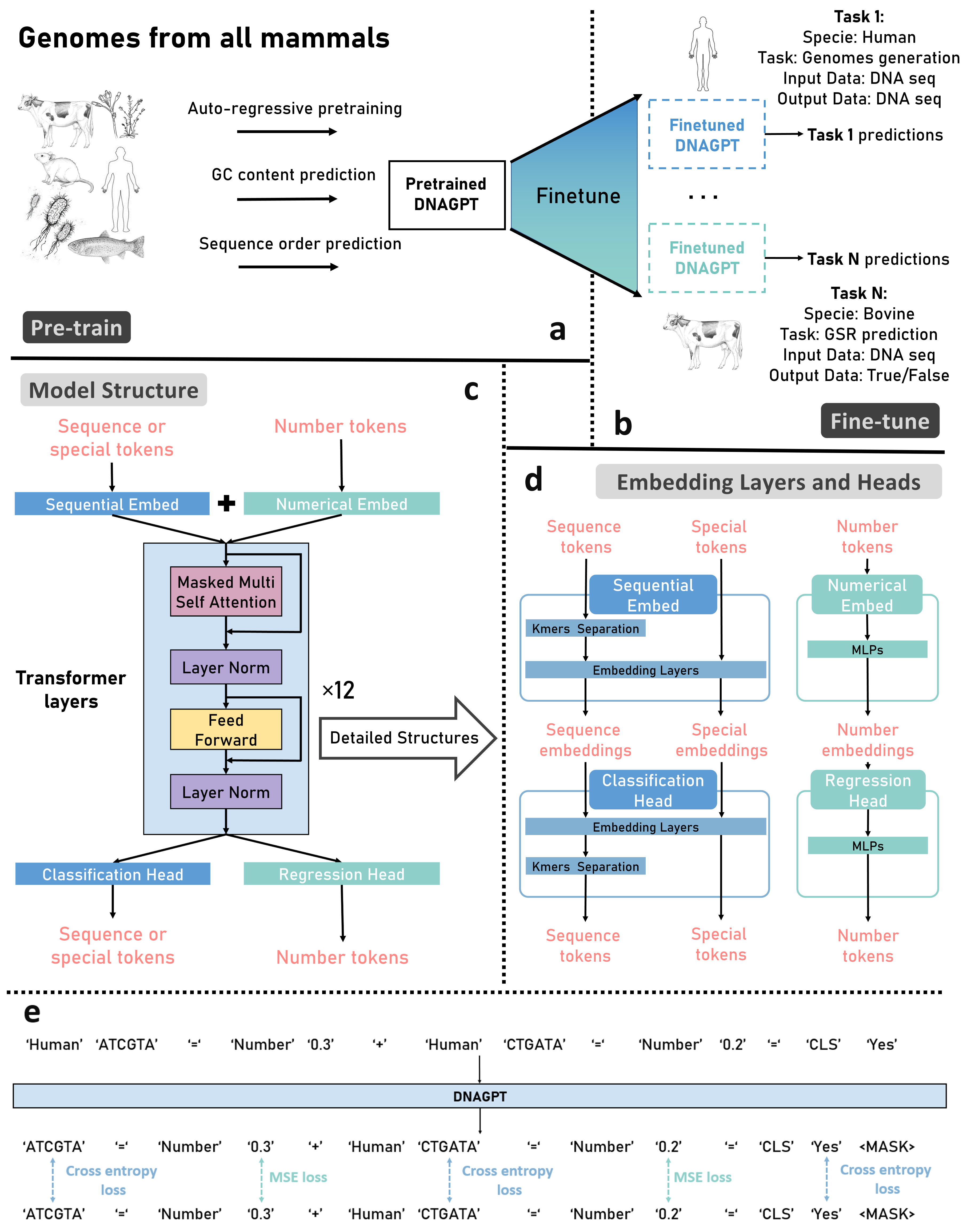}
\caption{Schematic of DNAGPT. \textbf{a}. The pre-training stage of DNAGPT. We utilize genomes from all mammals and design three pre-training tasks to jointly pre-train DNAGPT. \textbf{b}. The fine-tune stage of DNAGPT. After fine-tuning on the downstream task-related datasets, DNAGPT is able to handle specific tasks. Moreover, DNAGPT supports downstream tasks from different species, as well as various task and data formats. \textbf{c}. Model structure of DNAGPT. Different types of tokens are processed separately by different embedding layers, and then combined together as the input for the backbone. \textbf{d}. Details of the embedding layers and decoding heads. The figure illustrates the zoom-in view of different encoding heads. When processing the input data, we use different heads for mapping according to the data types. \textbf{e}. Model inputs (the first row), outputs (the second row) and ground truth (the third row) of DNAGPT. Tokens with different data types are evaluated with cross-entropy loss or mean squared error (MSE) loss. }\label{fig1:main}
\end{figure}

\section{DNAGPT architecture}
\subsection{Model structure}
The backbone of DNAGPT is a transformer-based \cite{vaswani2017attention} auto-regressive \cite{bollerslev1986generalized} decoder with the masked self-attention \cite{li2021mst} module. To better deal with numerical information, we pre-train the DNA sequence and numerical property end to end in a single model. The detailed network structure is presented in Figure. \ref{fig1:main} \textbf{c}. DNAGPT uses sequence tokens to denote the encoded DNA sequence and number tokens for the encoded numerical attributes. The sampled DNA sequence is first processed into a string of non-overlapped k-mers token input, then sent into the \textit{Sequential Embedding Layer} to be encoded as embeddings. The numbers are sent directly into a \textit{Numerical Embedding Layer} to be encoded as embeddings co-trained with the DNA embeddings. Then we concatenate both embeddings and send them into the GPT. The outputs of the GPT are split into two types of embeddings and sent to the \textit{Classification Head} to classify different tokens and \textit{Regression Head} to generate numbers, respectively. The structure of those heads is presented in Figure. \ref{fig1:main} \textbf{d}. It’s worth noting that DNAGPT can handle versatile downstream applications, where only fine-tuning of the original model parameters is needed. This simplifies the model’s usage, preserves its generalizability, and lays the foundation for potential zero-shot learning.

\begin{figure}[!t]%
\centering
\includegraphics[width=0.95\textwidth]{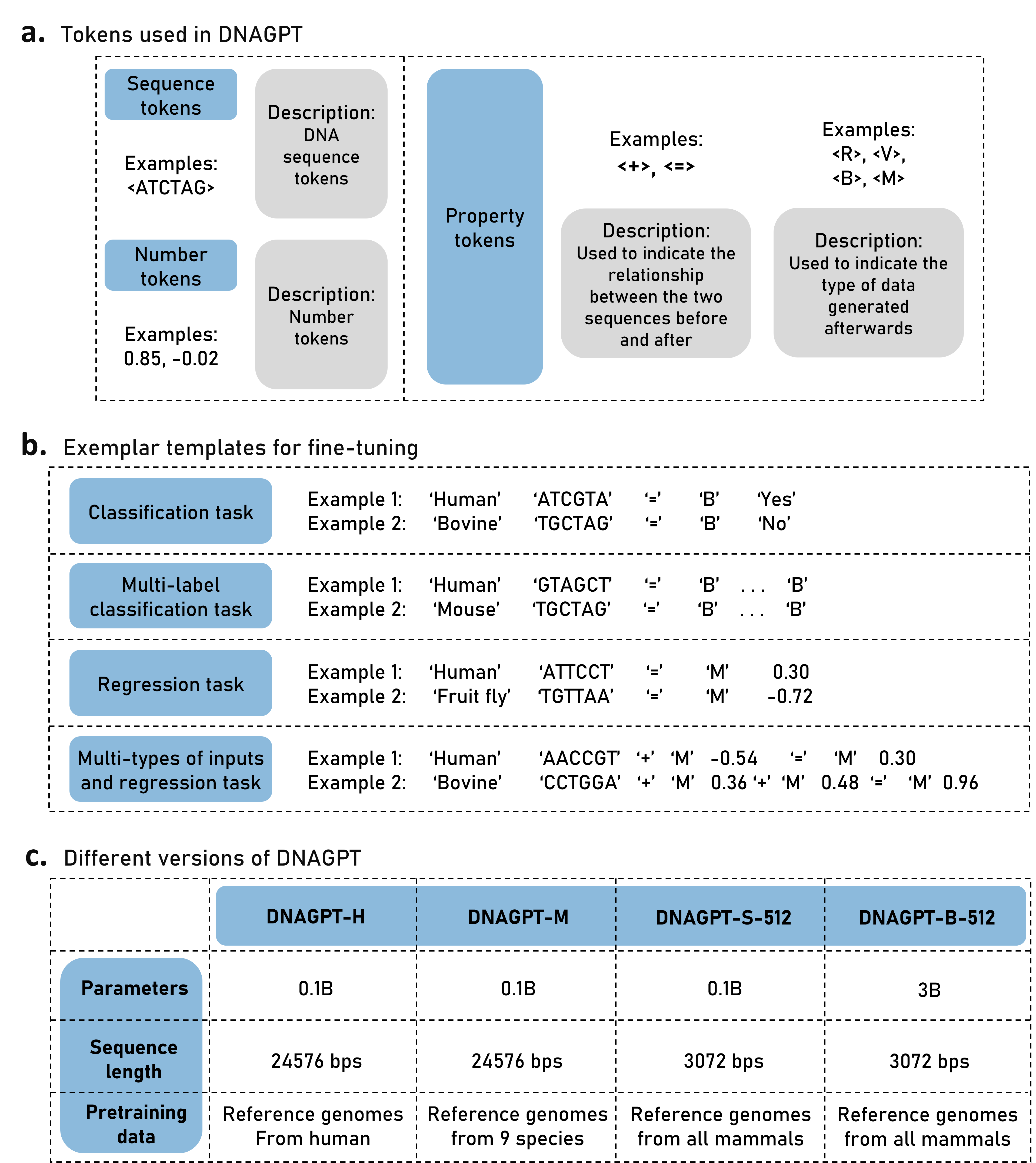}
\caption{Token language of DNAGPT. \textbf{a}. Tokens used in DNAGPT. \textbf{b}. Exemplar templates of the input and label in fine-tuning. \textbf{c}. Details of different versions of DNAGPT}\label{fig1:table}
\end{figure}


\subsection{Design of token language}

Currently, most DNA pre-training methods \cite{ji2021dnabert, dalla2023nucleotide} simply use strategies from natural language models and do not consider the characteristics of DNA sequence and specific biological tasks in the model design.
DNA sequence has no organizational structure as the nature language, which can be hierarchically divided into paragraphs, sentences, words and punctuations. 
We design a hierarchical token language structure for DNA sequences. Non-overlapped k-mers based on bps (base pairs) are first used to generate DNA words. DNA words of variable lengths are then combined to form DNA sentences. DNA sentences of varying lengths are then integrated to form DNA paragraphs, which are input into the GPT model.

As shown in Figure.  \ref{fig1:table} \textbf{a}, the regular input and output tokens are \textit{Sequence tokens} and \textit{Number tokens} which represent the DNA sequences and numbers respectively. Instruction tokens are used to give a prompt to the model about what are the next sequence of the tokens should the model output. Take an example, $'$Human$''$AATAAA$'$ indicates we encode a \textbf{human} AATAAA polyadenylation signals and $'$Bovine$''$AATAAA$'$ indicates we encode a \textbf{bovine} AATAAA polyadenylation signals. Similarly, $'$M$''$0.3155$'$ indicates that we encode a number into the model and in $'$B$''$X$'$, $'$B$'$ is the instruction token of the binary classification where the \textit{Classification tokens} $'$A$'$ indicates 'True' and $'$N$'$ indicates 'False'. Furthermore, to better construct connections, we use \textit{Connection tokens} to form the connections of two series of tokens, where $'$+$'$ represent the aggregation of two series of tokens and $'$=$'$ represent a relation of input and output. Specifically, when we want to predict the expression level of mRNA from both DNA sequence and the mRNA half-life values, we can encode the inputs as $'$Human$''$ATCGTC$''$+$''$M$''$-0.3484$''$=$''$M$''$0.9854$'$. This input indicates that we hope the model can generate the information from both of the $'$ATCGTC$'$ sequence and the input number $'$-0.3484$'$ to output the result numbers $'$0.9854$'$. The reserved tokens include numbers from $'$0$'$ to $'$9$'$, some unused uppercase letters like $'$K$'$, $'$L$'$, etc. and some special symbols like $'$*$'$ and $'$/$'$, etc. These reserved tokens can be used to build up more exclusive tasks for DNA sequence analysis. The complete token list is presented in the Figure. \ref{s_token}.

\section{Multi-tasks pre-training}

In order to integrate DNA sequence information from multiple species and allow downstream tasks to benefit from cross-species information, we proposed four variations of DNAGPT, named DNAGPT-H, DNAGPT-M, DNAGPT-S-512 and DNAGPT-B-512. As shown in Figure. \ref{fig1:table} \textbf{c}, DNAGPT-H, DNAGPT-M and DNAGPT-S-512 have 0.1 billion parameters and DNAGPT-B-512 has 3 billion parameters. Specifically, DNAGPT-H's sequence length is set to 4096, equivalent to 24,576 bps, and its pre-training data is based on Human reference genomes; DNAGPT-M also has a sequence length of 4096, with pre-training data from reference genomes of 9 species; DNAGPT-S-512 and DNAGPT-B-512 have a sequence length set to 512 and its pre-training data consists of reference genomes from all mammals. Specifically, the dataset for Genomes from 9 species includes reference genomes from \textit{Arabidopsis\_thaliana, Caenorhabditis\_elegans, Bos\_taurus, Danio\_rerio, Drosophila\_melanogaster, Escherichia\_coli\_gca\_001721525, Homo\_sapiens, Mus\_musculus, Saccharomyces\_cerevisiae} with a total of 10 billion bps. For the mammals' dataset, we downloaded all mammalian reference genomes from the NCBI GenBank. After preprocessing, approximately 200 billion bps of data were sampled for pre-training. We then compare the three versions of DNAGPT in the ablation study and provide a detailed description of the data used in the supplementary materials. Reported results in different tasks are from the suitable version of DNAGPT for each task due to the limitation of task-specific sequence length. In the GSR classification task, we used all three versions of DNAGPT. For the mRNA prediction and pseudo genomes generation tasks, the input sequence length requirements are greater than 512. Therefore, we utilize DNAGPTs with an input sequence length of 4096.

\subsection{Pre-training tasks}

We design three pre-training tasks for DNAGPT to fully characterize the DNA sequence and its associated numerical properties, including one standard GPT task and two DNA-specific tasks.

\paragraph{Next token prediction}
Next token prediction \cite{gillioz2020overview} is a classical pre-training task in NLP. GPT leverages this technique which can predict the next possible token based on the previous tokens. Recently, by adding more parameters and more training data, GPT-3 and GPT-4 demonstrate remarkable performance on various tasks. In DNAGPT, we also use the next token prediction strategy as the fundamental pre-training task.  

\paragraph{Guanine-cytosine content prediction}
Guanine-cytosine (GC) content plays a crucial role in transcriptome analysis as it provides essential information about genome structure, such as structural variations \cite{geoffroy2018annotsv} and transcriptional activity \cite{meyers2001abundance, dillon2015production}. In this task, we encode the GC content as number tokens in DNAGPT, allowing for joint training of numerical and sequence data and enabling DNAGPT to adapt to downstream tasks with numerical data as input and output. Furthermore, we adopt dynamic sequence length for the DNA sequence in this task, which allows the model to learn a dynamic receptive field and enables the downstream tasks with dynamic sequence length as input. We first calculate the GC content value of randomly selected sequences, which is an entirely unsupervised manner. The model should output this value after reading the entire sequence.

\paragraph{Sequence order prediction}

The sequence order of DNA plays an important role in gene expression \cite{basehoar2004identification} and transcription \cite{remmele2014transcriptional, korhonen2003twinkle}. For instance, sequences such as TATA box \cite{juo1996proteins} and AATAAA PAS \cite{mclauchlan1985consensus} often have to maintain a fixed order. We design a self-supervised sequence order prediction task, where we randomly reverse a sequence and let the model predict whether the sequence has been reversed or not. This task provides heuristic information for downstream tasks with order-sensitive sequences.

Since GPT models use unidirectional attention \cite{choromanski2020rethinking}, they can only infer and generate tokens from left to right. By reversing the DNA sequences, our model can infer tokens in both directions from the global perspective, improving its capability for downstream tasks for predicting preceding contexts.

\subsection{Pre-training Loss}
For the calculation of the loss in DNAGPT, as shown in Figure. \ref{fig1:main}. \textbf{e}, we illustrate the model input, output, and ground truth for DNAGPT during pre-training. The output of DNAGPT can be DNA tokens and/or number tokens. When calculating the loss for the next token prediction and sequence order prediction task, cross-entropy loss is used. For the GC ratio prediction task, mean squared error (MSE) loss is used since numerical tokens are involved. The final loss can be represented as:

\begin{equation}
    Loss = \lambda \times MSE\_loss + Cross\_entropy\_loss
\end{equation}
where $MSE\_loss$ indicates MSE loss and $Cross\_entropy\_loss$ indicates Cross entropy loss. In the pre-training, the $\lambda$ is set to 0.01.

\section{Genomic signals and regions (GSR) recognition}

Recognition of various genomic signals and regions (GSR) from DNA sequence is essential to the understanding of genomes. To address this issue, we fine-tune and evaluate our model on the recognition of polyadenylation signals (PAS) and translation initiation sites (TIS) of different organisms: human, mouse, bovine and fruit fly. To be specific, we follow the processing procedure in DeepGSR \cite{kalkatawi2019deepgsr}. The DNA sequence lengths are set to 603 and 606 respectively for TIS and PAS recognition. DeepGSR extracted 20,933, 18,693, 12,082, and 27,203 true PAS data; and 28,244, 25,205, 17,558, and 30,283 true TIS for human, mouse, bovine, and fruit fly, respectively which are used as groud-truth. Then Deepgsr sampled a similar number of non-GSR sequences from the genome sequences and combined them with the true cases. The training set, validation set, and test set are divided in the ratio of 6:1.5:2.5. Details of the datasets are depicted in Section \ref{dataset}. We report the results of DNAGPT-B-512 in this task.

\subsection{DNAGPT is able of recognizing GSRs from any species.}
\begin{figure*}[!]%
\centering
\includegraphics[width=0.95\textwidth]{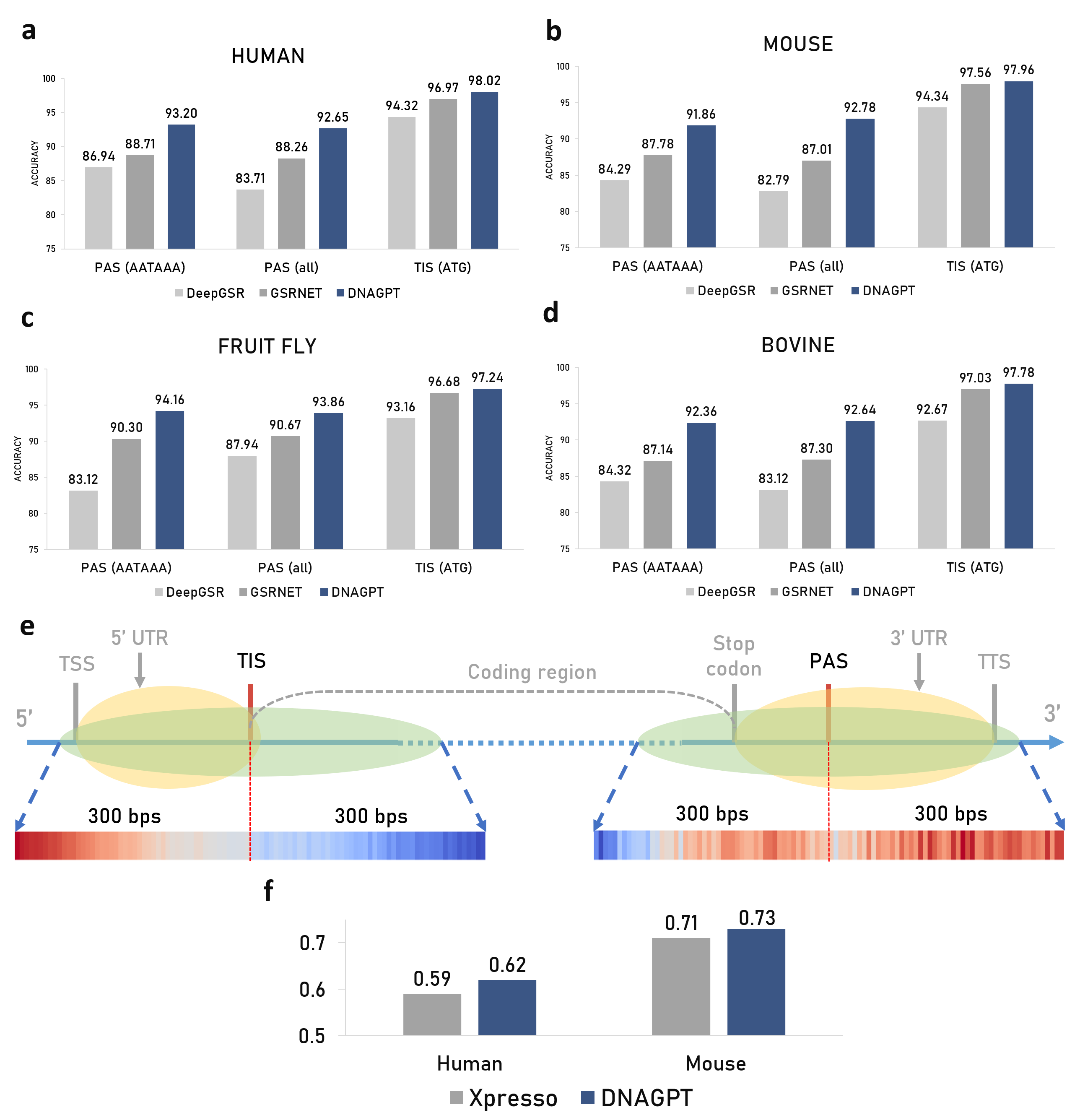}
\caption{Performance comparison between DNAGPT and other methods on PAS and TIS recognition. We fine-tune and evaluate our DNAGPTs on the data from four organisms, including human (\textbf{a}), mouse (\textbf{b}), fruit fly (\textbf{c}) and bovine (\textbf{d}). In each subgraph, we compare the accuracy of the DNAGPT with previous methods on the recognition of PAS (AATAAA), PAS (all) and TIS (ATG) from left to right. The reported results come from DNAGPT-B-512.\textbf{e}. Attention maps of the final layer of DNAGPT (last row). The green regions are the sampled input sequences and the GSRs are located in the middle of the sequence. The yellow regions are the 5' Untranslated Region and 3' Untranslated Region, respectively. 300 bps before and after the GSRs are included in the analysis. \textbf{f}. Performance comparison for DNAGPTs and 
other methods of mRNA abundance prediction. The reported $r^2$ results show that compared to mouse genomes, tasks on human genomes can benefit more by utilizing DNAGPT. The reported results come from DNAGPT-M.}
\label{classi}
\end{figure*}

The recognition of GSR can be considered as a binary classification task. We evaluate DNAGPT on the recognition of both PAS (AATAAA variant and all variants) and TIS (with the ATG signal) in the human genome. We present the accuracy metric in Figure. \ref{classi} \textbf{a}, which shows that our model can steadily outperform the previous state-of-the-art methods. We further provide additional metric results in the Table. \ref{s_result} and \ref{s_result_1} for a more comprehensive evaluation. 
Notice that, GSRNET \cite{zhu2023gsrnet} utilizes the embedded features generated from the pre-trained DNABERT model. DNAGPT can significantly outperform the modified DNABERT in all the tasks. To verify the generalization of DNAGPT, we further evaluate our model on other organisms, including mouse, fruit fly and bovine. Experimental results are presented in Figure. \ref{classi} \textbf{b}, \textbf{c} and \textbf{d}, respectively. Our DNAGPT outperforms the GSRNET and DeepGSR in most cases, the latter two were specially developed for GSR recognition. 


\subsection{DNAGPT recognizes GSRs based on non-coding regions.}

To explore the inner relations behind DNAGPT’s ability to recognize GSRs, we visualize the attention map of the final layer in DNAGPT’s backbone. The input data is TIS or PAS (AATAAA) sequence from humans, respectively. As shown in Figure. \ref{classi} \textbf{e}, we sample 300 bps before and after the TIS and PAS locations (green areas), which contain both coding and non-coding (yellow) regions. TIS is located right in front of the coding region, where is the non-coding region that DNAGPT focuses its attention and therefore accurately identifies TIS. DNAGPT shows the same attention pattern for the PAS recognition tasks. The attention maps of both cases adequately demonstrate that DNAGPT can recognize information in non-coding regions to identify GSRs.

\section{mRNA expression level prediction}

We then investigated whether DNAGPT could extract more abundant information from DNA sequences by attempting to predict the mRNA expression levels of corresponding promoters directly from genomic sequence information. Following Xpresso \cite{agarwal2020predicting}, we utilized 18,377 and 21,856 promoters as well as the mRNA half-lives in human and mouse respectively and held out 1000 cases in each specie for testing. CAP-Analysis Gene Expression (CAGE) was used to refine the annotations. Xpresso utilized deep convolutional network to encode both promoters and the half-lives and predicted the corresponding mRNA expression level and achieved much better results compared to traditional methods..

We used DNAGPT to predict the mRNA abundance under the same setting as Xpresso. We report the results of DNAGPT-M in this task. As mentioned in the last line of Figure. \ref{fig1:table} \textbf{b}. We combined the promoter sequences with the mRNA half-lives in a single sequence to predict the expression level of the mRNA abundance. We present the $r^2$ (Coefficient of determination) metric in Figure. \ref{classi} \textbf{f}. DNAGPT outperformed Xpresso from 0.59 to 0.62 for human mRNA abundance prediction and improved the results on the mouse species from 0.71 to approximately 0.73. 

The input format of this task where both sequence and numerical are provided can  not be handled by language-based models. Previously, specialized models such as Xpresso designed by experts have to be developed.  DNAGPT can handle these versatile tasks, obviating the need for designing more diverse and complex models.

\section{Artificial human genomes generation}

As the primitive task of the GPT model, we further investigate DNAGPT’s performance on the generation of artificial human genomes (AGs). AGs can be used to protect genetic privacy and reduce the cost of genetic sample collection. Following the work in \cite{yelmen2021creating}, we fine-tune our DNAGPT on 5008 haplotypes from 1000 Genomes data \cite{10002015global} which can be seen as the real genomes sequences and we use DNAGPT to generate 5000 AGs of 10000 Single Nucleotide Polymorphisms (SNPs) region for further analysis (can be seen as 5000 sequences each with a length of 10,000 bps). We compared DNAGPT with the GAN and RBM models. The GAN model consists of a generator and a discriminator network, where the output of the generator and the input of the discriminator both have the size of the number of SNPs. For the RBM model, we use the RBM model provided in \cite{yelmen2021creating}. All the training and testing strategy of GAN and RBM remains the same with \cite{yelmen2021creating}. We use the real 5008 haplotypes for the comparisons for all the methods (GAN, RBM, DNAGPT). We report the results of DNAGPT-M in this task.

\subsection{Analysis of artificial human genomes}

We evaluate DNAGPT and comparison methods from the following perspectives: principal components (PC) \cite{abdi2010principal}; allele frequency (AF) \cite{boehnke1991allele}, linkage disequilibrium (LD) \cite{reich2001linkage} and Pairwise haplotype distances. The evaluation metrics include Wasserstein distances \cite{arjovsky2017wasserstein} and correlation ($r^2$).

\begin{figure*}[]%
\centering
\includegraphics[width=1.0\textwidth]{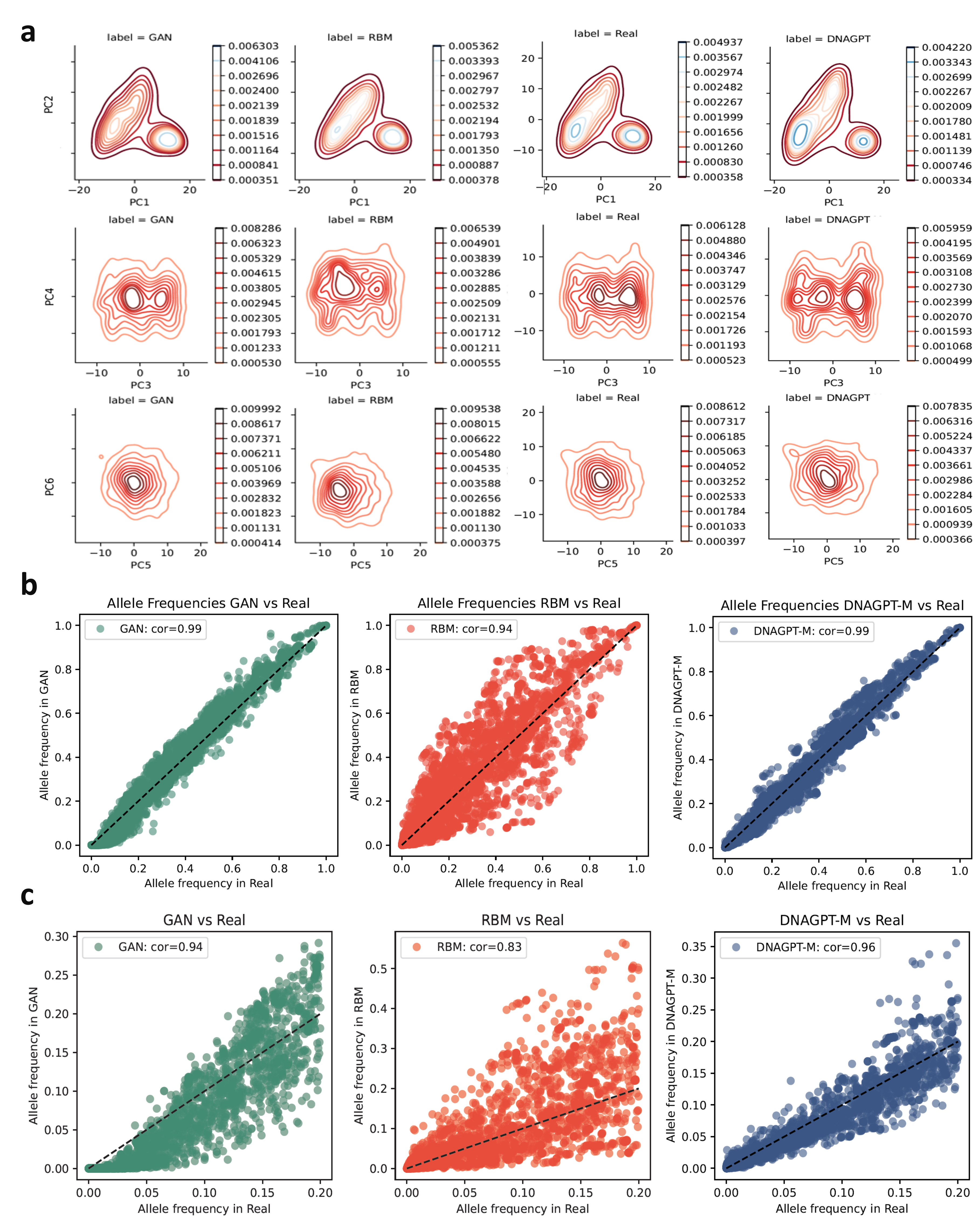}
\caption{
\textbf{a}. Comparison of isoline maps of PCA results of artificial human genomes generation. GAN; RBM; Real; DNAGPT.
\textbf{b}. Correlations of allele frequency between real genomes and artificial genomes. 
\textbf{c}. Correlations of allele frequency between real genomes and artificial genomes, specific on the sites with allele frequency less than 0.2 in the real genomes.
}\label{pca}
\end{figure*}

\paragraph{Principal components}

We conduct the principal component analysis (PCA) on the AGs generated from GAN, RBM, and DNAGPT. We show the value distribution of the first six principal components using an isoline map in Figure. \ref{pca} \textbf{a}. Results show that the distributions of AGs generated from all methods roughly align with those of the real human genomes, while DNAGPT model demonstrates the most similar distribution of the real sequences. We further compute the Wasserstein distance (lower is better) between distributions of AGs and real genome sequence, which are 1.753. 3.432, 1.131 for GAN, RBM, DNAGPT, respectively.

\paragraph{Allele frequency}

Allele frequency analysis is a genetic analysis method used to determine the frequency of different alleles of a gene locus.  The allele frequency at a polymorphic site depends on the variation of that site in all cases. In this analysis, we detect the frequency of SNPs within the 5,000 AGs from all the methods as well as the 5008 real AGs.  We conduct the analysis of the sequences generated by all the models. As shown in Figure. \ref{pca} \textbf{b}, both the DNAGPT and GAN perform stably  with a correlation of 0.99.  We then visualize the correlation of those sites with allele frequency less than 0.2. As shown in Figure. \ref{pca} \textbf{c},  DNAGPT outperforms GAN (0.94) and RBM (0.83) with a correlation of 0.96, indicating  that DNAGPT can better capture the information even from low-frequency alleles.

\begin{figure*}[!]%
\centering
\includegraphics[width=1.0\textwidth]{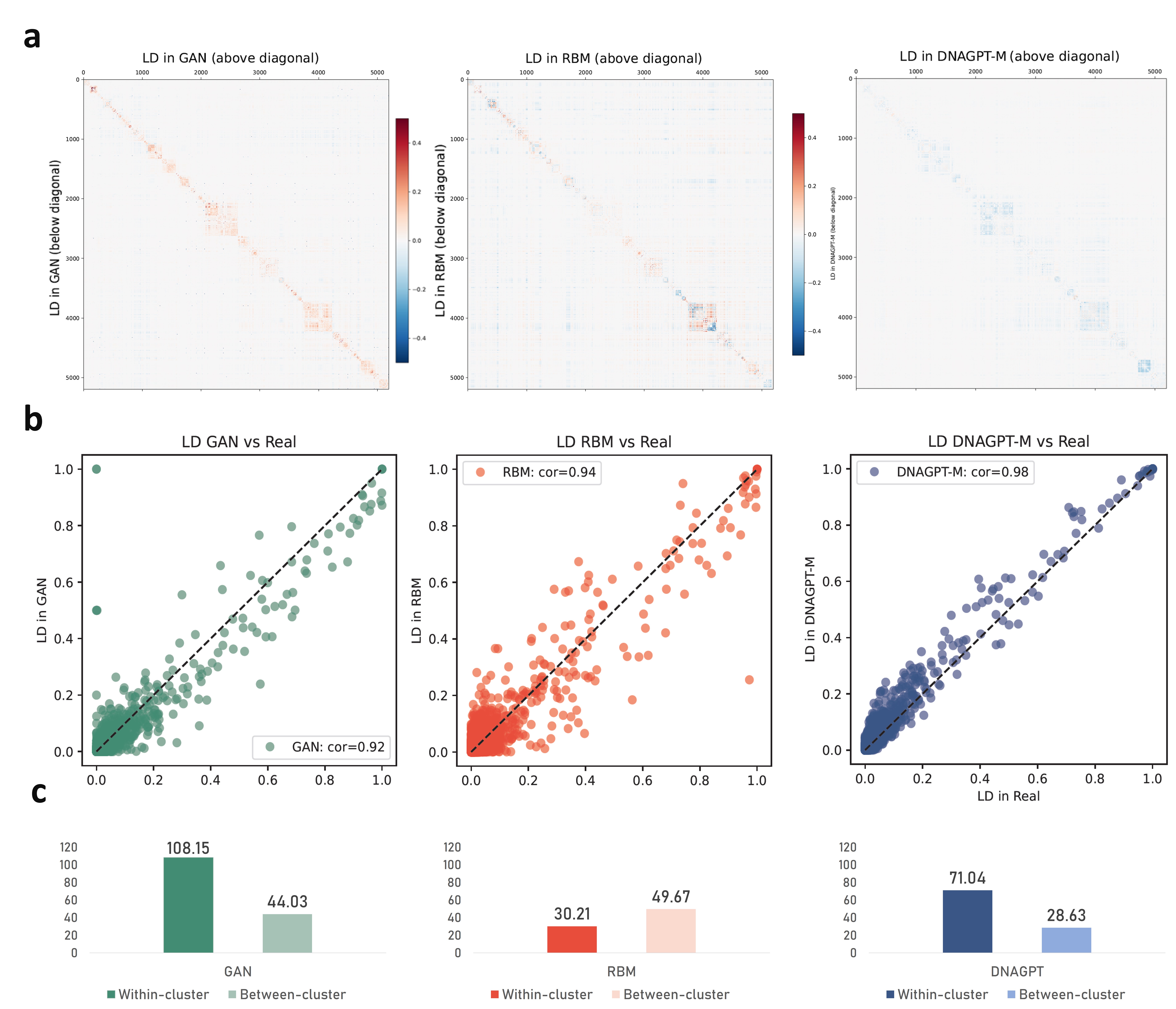}
\caption{ 
We placed the results of the same method in one column, specifically, the first column is GAN vs. real; the second column is RBM vs. real; the third column is DNAGPT vs. real. Each row represents:
\textbf{a}. Normalized correlation matrices of SNPs. We subtracted the correlation matrix of each method from the real genomes. The lighter the color, the closer the artificial genomes are to the real genomes.
\textbf{b}. Correlations of LD between real genomes and artificial genomes.
\textbf{c}. Wasserstein distances of pairwise haplotype distance distribution.
}\label{af}
\end{figure*}

\paragraph{Linkage disequilibrium}

Linkage disequilibrium (LD) is a phenomenon in population genetics that can be defined as the correlations of frequencies of two or more genetic markers (like alleles or genes). We further analyze the LD for all the generated sequences and real sequences. Figure. \ref{af} \textbf{a} illustrates the difference in LD values between human genomes generated by GAN, RBM and DNAGPT compared to real genomes, respectively. In these panels, the lighter the color, the more similar the LD heat map is to the real genomes. Among them, the LD of DNAGPT is slightly weaker than that of real genomes, while GAN and RBM are stronger than the original genomes. Overall, the heat map performance of DNAGPT is better than GAN and RBM, as their colors are lighter. The above conclusions can also be verified through a comparison of correlation values. We present the correlation distributions in Figure. \ref{af} \textbf{b}. The correlation between the LDs of real and generated sequences from GAN and RBM is 0.92 and 0.94 and DNAGPT can achieve a score of 0.98.


\paragraph{Pairwise haplotype distances analysis}

Pairwise haplotype distances refer to the genetic distances between different haplotypes within a genome. When calculating the distances, we typically compare the differences in the alleles at the corresponding loci between two haplotypes. In this analysis, we first calculate the pairwise distance distributions within each cluster of generated genomes (GAN vs GAN, RBM vs RBM, DNAGPT vs DNAGPT), defined as \textit{Within-cluster}, then the pairwise distance distributions between real genomes and generated genomes by each method (GAN vs Real, RBM vs Real, DNAGPT vs Real) are defined as \textit{Between-cluster}. Then we calculate the Wasserstein distances between the two types of distributions within the distribution of real genomes (Real vs Real). We present the Wasserstein distances of within-cluster in Figure. \ref{af} \textbf{c}. Among them, the GAN’s distribution has the largest gap compared to the actual distribution with a value of 108.15, followed by DNAGPT with a value of 71.04. The genomes generated by RBM have the smallest discrepancy with a value of 30.21 from real genomes. The Between-cluster reflects the discrepancy between the pairwise distance distribution of genomes generated by each method and real genomes. The genomes generated by DNAGPT are the most similar to the real genomes with a value of 28.63, while RBM performs the worst, followed closely by GAN.

\subsection{Generation temperature of DNAGPT can influence the quality of generated genomes}

\begin{figure}[!]%
\centering
\includegraphics[width=0.95\textwidth]{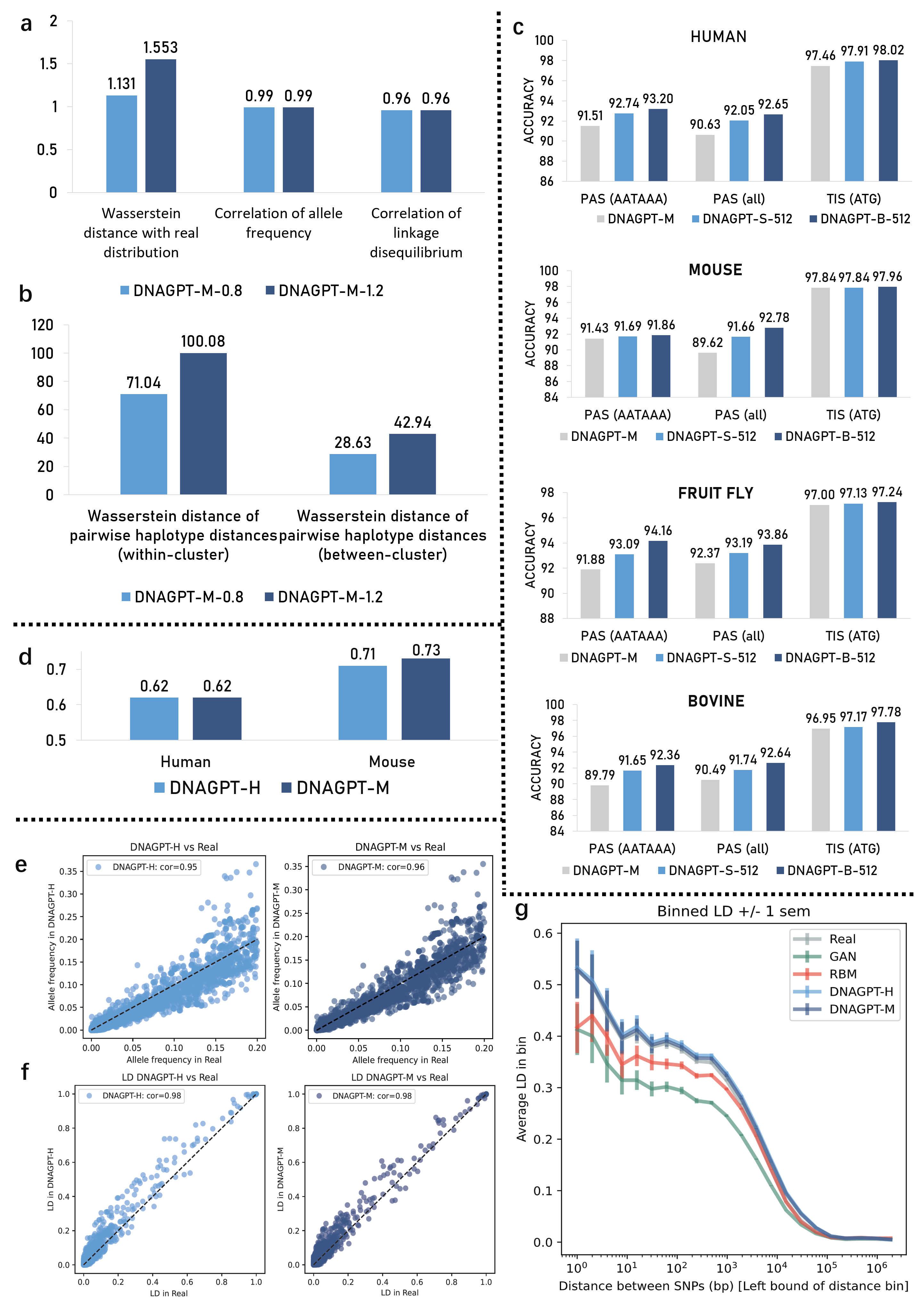}
\caption{
\textbf{a}. Comparisons of Wasserstein distance, Correlation of allele frequency, and Correlation of linkage disequilibrium for DNAGPTs with generation temperature 0.8 and 1.2 respectively.
\textbf{b}. Comparisons of Wasserstein distance of pairwise haplotype distance distribution for DNAGPTs with generation temperature 0.8 and 1.2 respectively.
\textbf{c}. Performance comparison for different DNAGPT on GSR recognition tasks.
\textbf{d}. Performance comparison for different DNAGPT on mRNA abundance prediction tasks.
\textbf{e}. Correlations of allele frequency between genomes generated by DNAGPT-H and DNAGPT-M, specific on the sites with allele frequency less than 0.2 in the real genomes.
\textbf{f}. Correlations of LD between genomes generated by DNAGPT-H and DNAGPT-M.
\textbf{g}. Average LD as a function of SNP distance after removing sites that are fixed in at least in one dataset. Pairwise SNP distances were stratified into 50 bins and for each distance bin, the correlation was averaged over all pairs of SNPs belonging to the bin. Green: GAN; Red: RBM; Light blue: DNAGPT-H; Dark blue: DNAGPT-M.
}\label{regre}
\end{figure}

When a trained DNAGPT generates the DNA sequence, we can control the randomness of the output sequence by adjusting the \textit{generation temperature}. The generation temperature ranges from 0 to infinity. The higher the generation temperature, the more random the generated sequence will be. In the experiments mentioned earlier, our default generation temperature was 0.8. In this section, we will adjust the generation temperature to 1.2 to evaluate the performance of DNAGPT under different generation temperatures. The results are shown in the Figure. \ref{regre} \textbf{a} and \textbf{b}. Figure. \ref{regre} \textbf{a} shows the Wasserstein distance, correlations of allele frequency, and correlations of linkage disequilibrium with the real distribution. Figure. \ref{regre} \textbf{b} shows the Wasserstein distance of pairwise haplotype distance distribution (within-cluster and between-cluster). We can find that a larger generation temperature allows DNAGPT to maintain the correlation of allele frequency and linkage disequilibrium virtually unchanged while increasing the distance from the real distribution. It also increases the Wasserstein distance of pairwise haplotype distance distribution, indicating that a larger generation temperature makes the generated DNA sequences more diverse, and the gap from the original distribution will slightly increase. Therefore, users can adjust the generation temperature according to their needs, thereby controlling the diversity and authenticity of the generated sequences.

\section{Comparisons of different versions of DNAGPT}

In this section, we compared the results of three different DNAGPT variations. We conducted comparisons in GSR prediction, mRNA expression level prediction, and artificial human genomes generation task. We report the results in Figure. \ref{regre}.
In the GSR prediction task, we compared the three different DNAGPT variations in Figure. \ref{regre} \textbf{c}. It can be seen that as the amount of pre-training data increases (Human reference genomes - reference genomes from 9 species - reference genomes from all mammals), the performance of downstream tasks also improves. This phenomenon can also be observed in the mRNA expression level prediction task. In the Figure. \ref{regre} \textbf{d}, although DNAGPT-M and DNAGPT-H are neck-and-neck in the human mRNA expression level prediction task, DNAGPT-M performs better than DNAGPT-H in the mouse mRNA expression level prediction task. 

We further compared DNAGPT-H and DNAGPT-M in the artificial human genomes generation task. In the Figure. \ref{regre} \textbf{e}, the correlations of allele frequency for the genomes generated by DNAGPT-M and DNAGPT-H are almost the same, with DNAGPT-M being slightly better at 0.96 compared to DNAGPT-H at 0.95. For the Correlations of LD of genomes, as can be seen from the Figure. \ref{regre} \textbf{f}, both DNAGPT-M and DNAGPT-H maintain an excellent level with a value of 0.98. From this, we further investigated the performance level of LD when considering different distances between SNPs. The Figure. \ref{regre} \textbf{g} shows that both DNAGPT variations fit the real data distribution better than GAN and RBM, with DNAGPT-M being slightly better than DNAGPT-H.

\section{Discussion}

In summary, we have developed a multi-task pre-training model called DNAGPT for DNA sequence analysis to accommodate versatile downstream tasks across multiple species. We conducted the pre-training on reference genomes from as many as 9 different species. Meanwhile, we introduced joint training of numbers and sequences during the pre-training process. In order to better encode the relationships between inputs and outputs for versatile task formats, we designed a set of token languages to incorporate sequence, number, and control tokens. For the pre-training tasks, to better understand the uniqueness of DNA sequences and the next token prediction task in GPT, we also introduced two pre-training tasks: GC content prediction and sequence order prediction. Finally, we utilized the token language to compile mixed inputs and outputs of DNA sequences and numerical properties. 

Our evaluation of DNAGPT on genomic signals and regions recognition tasks showed that the model can accurately determine whether a given DNA sequence is a genuine genomic signal or region. Furthermore, DNAGPT can also handle joint inputs of DNA sequences and mRNA half-lives to predict mRNA expression levels. In the Artificial human genomes generation task, the AGs generated by DNAGPT rank highly in various evaluation metrics, indicating that DNAGPT effectively comprehends the underlying relationships and information within genomes.

Despite its promising results, DNAGPT has several limitations that warrant further investigation. One such limitation is the model's current focus on DNA sequences. Extending DNAGPT to handle multi-omics and spatial-omics data would greatly enhance its applicability and enable more comprehensive analyses of biological tasks. Another area for improvement is the incorporation of multi-modal data, such as pathology tissue images and disease diagnostic reports, which would provide a more holistic perspective on biological tasks. Additionally, addressing the challenge of processing long sequence data, which is common in biological research, could be achieved by employing memory-efficient model structures, such as RWKV \cite{peng2023rwkv} and RetNet \cite{sun2023retentive}.

Finally, the efficient adaptation of DNAGPT should be explored, as users may not have the resources to fine-tune the model. Techniques for efficient training of foundation models \cite{hu2021lora, hu2023llm, wang2023bootstrap} could be tested and developed with DNAGPT, and zero-shot adaptation or 'emergent' abilities in biology foundation models could be further investigated. Despite these limitations, exploring new avenues for DNAGPT's development holds the potential to significantly advance our understanding of DNA sequences and contribute to a wide array of biological research applications.

\section{Methods}

\paragraph{Pre-training of DNAGPT}

For DNAGPT-H, we collect the reference genomes from the Ensembl database \cite{cunningham2022ensembl} with a total amount of 3 billion bps. During the data sampling stage, we employed a non-overlapped k-mers sampling strategy to handle DNA sequence data. While sampling, we removed sequences with an 'N'(denoted as "not detected") content ratio greater than 0.05. Moreover, we performed random flipping with a probability of 0.5. we then encoded each input DNA sequence and numerical information according to the token language and the pre-training tasks we designed. DNAGPT-H consists of 12 layers of transformer blocks based on unidirectional attention, with each layer containing 12 attention heads and a hidden layer size of 768. The number of trained parameters in the model is 0.1 billion. The learning rate is set to 1e-4 with a cosine decay scheduler. The weight decay is set to 1e-2. The optimizer we choose is AdamW with the betas set to (0.9, 0.95) and momentum set to 0.937. We employed mixed precision for pre-training. The model was pre-trained for 15 epochs. The pre-training of the model on 8 Nvidia V100 32GB GPUs took approximately one day.

For DNAGPT-M, we collected reference genome information of 9 species from the Ensembl database \cite{cunningham2022ensembl}, including \textit{arabidopsis\_thaliana, caenorhabditis\_elegans, bos\_taurus, danio\_rerio, drosophila\_melanogaster, escherichia\_coli\_gca\_001721525, homo\_sapiens, mus\_musculus, saccharomyces\_cerevisiae}. Subsequently, we removed the mitochondrial genomes from the majority of the species in the preprocessing procedure. After preprocessing, the number of bps in the genome of each species is: arabidopsis\_thaliana (119146348 bps), caenorhabditis\_elegans (100272607 bps), bos\_taurus (2628394923 bps), danio\_rerio (1345101833 bps), drosophila\_melanogaster (137547960 bps), escherichia\_coli\_gca\_001721525 (5176750 bps), homo\_sapiens (3088286401 bps), mus\_musculus (2723414844 bps), saccharomyces\_cerevisiae (12071326 bps). The total amount of bps is 10159412992. The architecture and training strategies are the same as DNAGPT-H. 

Similar to DNAGPT-M, DNAGPT-S-512 used the same model as well as the hyperparameters, but the pre-training data changed from genomes of 9 species to the reference genomes of all the mammals with a total amount of approximately 200 billion bps. DNAGPT-S-512 was trained on the data for 2 epochs and took approximately one week to finish the pre-training stage.

\paragraph{Non-overlapping k-mers tokenization}

A k-mer strategy composes k consecutive nucleotides into one token. Previous k-mers methods often adopt overlapped tokenization, that is, regardless of the value of k, the shift during each sampling is always $1$, resulting in $(N+k-1)$ tokens for a N-length sequence. In the non-overlapped k-mers strategy, the shift is equal to K, resulting in $N/k$ tokens for an N-length sequence and improving the efficiency by k times. 


\paragraph{Fine-tuning of DNAGPT}

When fine-tuning DNAGPTs, Firstly, we should set the input sequence information to organize the data and initialize the model, and the model can automatically initialize suitable encoding heads. For example, for classification and generation tasks, the sequence embedding and classification heads are activated for input and output. For regression tasks and more complex composite tasks, DNAGPT first composes the input for joint embeddings and then selects regression heads for task output. After the embedding layer and task heads are set, the pre-trained weights are loaded into the model, and the weights of unused heads will be discarded. Then we can fine-tune DNAGPTs using data from the downstream tasks. We use the same hyperparameters across all downstream tasks. For 0.1B models, the hyperparameters are set to: max learning rate, $3 \times 10^{-5}$; learning scheduler, cosine with warmup; optimizer, AdamW; warmup epoch, 3; weight decay, $1e-1$; batch size, 8; For 3B models, the hyperparameters are set to: max learning rate, $3 \times 10^{-6}$; learning scheduler, cosine with warmup; optimizer, AdamW; warmup epoch, 3; weight decay, $1e-1$; batch size, 8.

In the genomic signals and region recognition, we use the sequence embedding and classification head. The evaluation metrics are ACC (Accuracy), F1 (F1 score), MCC (Matthews Correlation Coefficient), Precision, and Recall. We report the complete results in the Table. \ref{s_result}. In mRNA expression levels prediction, both the sequence embedding and the number embedding are invoked to handle the input of sequences and numbers. For the output, the regression head is used to predict the expression level. In artificial human genomes generation, only the sequence embedding and classification head are used to handle input and output sequences. During fine-tuning, we add a stop symbol at the last position of the input sequence. When generating sequences, we remove all sequences that do not have the stop symbol or those with incorrect stop symbol positions in the post-processing step. For temperature adjustment, we keep the training epoch and other hyper-parameters unchanged.

\bibliography{sn-bibliography}

\clearpage

\setcounter{section}{0}
\setcounter{figure}{0}
\setcounter{table}{0}
\setcounter{equation}{0}
\renewcommand{\thesection}{S\arabic{section}}
\renewcommand{\thefigure}{S\arabic{figure}}
\renewcommand{\thetable}{S\arabic{table}}
\renewcommand{\theequation}{S\arabic{equation}}

\section{Supplementary}

\subsection{Comparisons to other models}
We further compare the performance on the datasets used in NT \cite{dalla2023nucleotide}, this dataset contains more GSR recognition tasks.All the tasks in the dataset are classification tasks. For DNAGPT-S-512, the hyperparameters are set to: max learning rate, $3 \times 10^{-5}$; learning scheduler, cosine with warmup; optimizer, AdamW; warmup epoch, 3; weight decay, $1e-1$; For DNAGPT-B-512, the hyperparameters are set to: max learning rate, $3 \times 10^{-6}$; learning scheduler, cosine with warmup; optimizer, AdamW; warmup epoch, 3; weight decay, $1e-1$.
The results are proposed in \ref{nt_result}. Our DNAGPT-B-512 is comparable to the NT-2.5B-850 model, and DNAGPT-S-512 is comparable to the NT-2.5B-3202 model in the NT dataset.

\begin{table}[h]
\caption{Full results of DNAGPT-B-512 on NT datasets. The Matthews correlation coefficient (MCC) is used as the metric.}
\begin{tabular}{@{}lccccc@{}}
\toprule
Task name      & \makecell[c]{NT\\500M-1} & \makecell[c]{NT\\2.5B-3202} & \makecell[c]{NT\\2.5B-850} & DNAGPT-S-512 & DNAGPT-B-512 \\
\midrule
H3             & 72.0   & 75.0  & 79.0   & 75.0         & \textbf{81.0}      \\
H3K4me1        & 36.0   & 42.0  & \textbf{54.0}   & 41.0         & 53.0      \\
H3K4me2        & 27.0   & 28.0  & \textbf{32.0}   & 26.0         & \textbf{32.0}      \\
H3K4me3        & 24.0   & 31.0  & \textbf{41.0}   & 32.0         & 38.0      \\
H3K9ac         & 45.0   & 49.0  & 55.0   & 48.0         & \textbf{56.0}      \\
H3K14ac        & 37.0   & 45.0  & \textbf{54.0}   & 46.0         & 52.0      \\
HK36me3        & 45.0   & 53.0  & \textbf{62.0}   & 56.0         & 58.0      \\
HK79me3        & 57.0   & 57.0  & \textbf{62.0}  & 57.0         & 61.0      \\
H4             & 75.0   & 79.0  & 81.0   & 78.0         & \textbf{83.0}      \\
H4ac           & 33.0   & 41.0  & \textbf{49.0}   & 43.0         & 47.0      \\
Promoter all          & 88.0   & 91.0  & 91.0   & 91.0         & \textbf{93.0}      \\
Promoter non-tata     & 91.0   & 93.0  & 94.0   & 92.0         & \textbf{95.0}       \\
Promoter tata         & 78.0   & 76.0  & 79.0   & 80.0         & \textbf{83.0}       \\
\botrule
\end{tabular}
\label{nt_result}
\end{table}

\subsection{Other results of DNAGPTs on genomic signals and regions recognition}

\paragraph{Full results of DNAGPTs on genomic signals and regions recognition}
We show in the Table. \ref{s_result} the results of DNAGPT-M on various datasets of GSR recognition task, and the results of DNAGPT-S-512 in the Table. \ref{s_result_1}. Bothe of the DNAGPTs demonstrates stable results across different GSR recognition datasets from various species and the performance of DNAGPT-S-512 is the best..

\begin{table}[h]
\caption{Full results of DNAGPT-M on genomic signals and regions recognition.}
\begin{tabular}{@{}lccccc@{}}
\toprule
Task name      & acc(\%) & f1(\%) & mcc(\%) & precision(\%) & recall(\%) \\
\midrule
Human\_PAS(AATAAA)    & 91.51   & 91.51  & 82.99   & 91.52         & 91.47      \\
Human\_PAS(all)       & 90.63   & 90.64  & 81.28   & 90.64         & 90.64      \\
Human\_TIS(ATG)       & 97.46   & 97.46  & 94.92   & 97.47         & 97.46      \\
Mouse\_PAS(AATAAA)    & 91.43   & 91.41  & 82.83   & 91.40         & 91.43      \\
Mouse\_PAS(all)       & 89.62   & 89.62  & 79.24   & 89.63         & 89.61      \\
Mouse\_TIS(ATG)       & 97.84   & 97.84  & 95.68   & 97.85         & 97.83      \\
Fruitfly\_PAS(AATAAA) & 91.88   & 91.87  & 83.84   & 91.96         & 91.88      \\
Fruitfly\_PAS(all)    & 92.37   & 92.38  & 84.76   & 92.38         & 92.38      \\
Fruitfly\_TIS(ATG)    & 97.00   & 97.00  & 94.01   & 97.00         & 97.00      \\
Bovine\_PAS(AATAAA)   & 89.79   & 89.77  & 79.65   & 89.89         & 89.76      \\
Bovine\_PAS(all)      & 90.49   & 90.49  & 80.99   & 90.49         & 90.49      \\
Bovine\_TIS(ATG)      & 96.95   & 96.95  & 93.90   & 96.95         & 96.95      \\
\botrule
\end{tabular}
\label{s_result}
\end{table}

\begin{table}[h]
\caption{Full results of DNAGPT-S-512 on genomic signals and regions recognition.}
\begin{tabular}{@{}lccccc@{}}
\toprule
Task name      & acc(\%) & f1(\%) & mcc(\%) & precision(\%) & recall(\%) \\
\midrule
Human\_PAS(AATAAA)    & 92.74   & 92.74  & 85.49   & 92.75         & 92.74      \\
Human\_PAS(all)       & 92.05   & 92.04  & 84.11   & 92.07         & 92.04      \\
Human\_TIS(ATG)       & 97.91   & 97.91  & 95.83   & 97.92         & 97.95      \\
Mouse\_PAS(AATAAA)    & 91.69   & 91.69  & 83.39   & 91.69         & 91.70      \\
Mouse\_PAS(all)       & 91.66   & 91.66  & 83.33   & 91.66         & 91.67      \\
Mouse\_TIS(ATG)       & 97.84   & 97.84  & 95.79   & 97.85         & 97.85      \\
Fruitfly\_PAS(AATAAA) & 93.09   & 93.09  & 86.17   & 93.08         & 93.08      \\
Fruitfly\_PAS(all)    & 93.19   & 93.18  & 86.47   & 93.27         & 93.19      \\
Fruitfly\_TIS(ATG)    & 97.13   & 97.13  & 94.28   & 97.14         & 97.13      \\
Bovine\_PAS(AATAAA)   & 91.65   & 91.65  & 83.31   & 91.68         & 91.64      \\
Bovine\_PAS(all)      & 91.74   & 91.74  & 83.50   & 91.75         & 91.74      \\
Bovine\_TIS(ATG)      & 97.17   & 97.17  & 94.34   & 97.17         & 97.16      \\
\botrule
\end{tabular}
\label{s_result_1}
\end{table}

\begin{table}[h]
\caption{Full results of DNAGPT-B-512 on genomic signals and regions recognition.}
\begin{tabular}{@{}lccccc@{}}
\toprule
Task name      & acc(\%) & f1(\%) & mcc(\%) & precision(\%) & recall(\%) \\
\midrule
Human\_PAS(AATAAA)    & 93.20   & 93.20  & 86.73   & 93.20         & 93.20      \\
Human\_PAS(all)       & 92.65   & 92.65  & 85.02   & 92.66         & 92.65      \\
Human\_TIS(ATG)       & 98.02   & 98.02  & 96.30   & 98.02         & 98.02      \\
Mouse\_PAS(AATAAA)    & 91.86   & 91.86  & 84.22   & 91.87         & 91.87      \\
Mouse\_PAS(all)       & 92.78   & 92.78  & 85.08   & 92.79         & 92.79      \\
Mouse\_TIS(ATG)       & 97.96   & 97.96  & 95.93   & 97.96         & 97.96      \\
Fruitfly\_PAS(AATAAA) & 94.16   & 94.16  & 87.78   & 94.16         & 94.16      \\
Fruitfly\_PAS(all)    & 93.86   & 93.86  & 87.14   & 93.86         & 93.86      \\
Fruitfly\_TIS(ATG)    & 97.24   & 97.24  & 94.28   & 97.24         & 97.24      \\
Bovine\_PAS(AATAAA)   & 92.36   & 92.36  & 84.68   & 92.36         & 92.37      \\
Bovine\_PAS(all)      & 92.64   & 92.64  & 84.92   & 92.64         & 92.64      \\
Bovine\_TIS(ATG)      & 97.78   & 97.78  & 94.92   & 97.78         & 97.78       \\
\botrule
\end{tabular}
\label{s_result_2}
\end{table}

\begin{figure*}[h!]%
\centering
\includegraphics[width=1.0\textwidth]{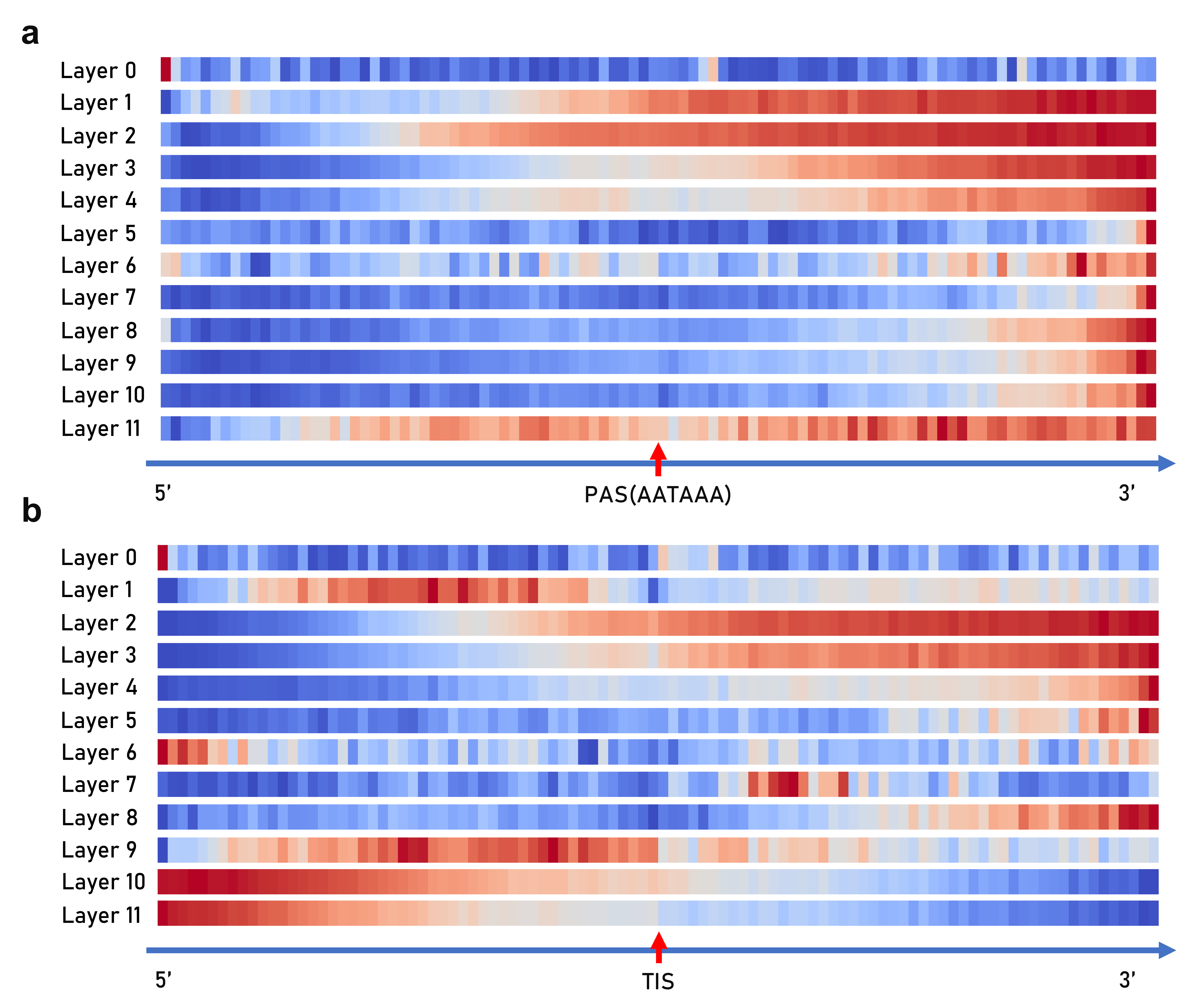}
\caption{Attention maps of each layer of DNAGPT-M with TIS input.}
\label{s_tis}
\end{figure*}

\paragraph{Attention maps of DNAGPT-M}

We show the attention map of each layer in DNAGPT-M in Figure \ref{s_tis} \textit{a}. The input sequence is PAS (AATAAA) sequence where the PAS site is located in the middle of the sequence. We can observe that almost all layers focus on the latter half of the area, with shallow and deep layers having a more widespread attention compared to the middle layers. We can also notice that the attention map of the shallow areas is smoother than that of the deep areas. Although the attention range of the deep layers is as extensive as those of the shallow layers, the deep networks tend to focus on a few specific tokens rather than presenting a smooth state like the shallow attention map. This indicates that some regions in non-coding areas may be more critical for PAS recognition compared to other areas. We have also displayed the attention map for each layer with TIS data. In the Figure. \ref{s_tis} \textit{b}, we display the attention maps of each layer of DNAGPT-M with TIS input. Interestingly, compared to the attention map with PAS as input, the information focused on by the model in the shallow layers is more consistent, with a notable difference only in Layer 1. In the later layers, the attention map for TIS input starts to focus on information from tokens in earlier positions, i.e., non-coding region information. This suggests that the information the model focuses on in the shallow layers is more approximate, but in the deep networks, it can more precisely pinpoint the locations of important tokens.

\subsection{All tokens used in DNAGPT}

\begin{figure*}[h!]%
\centering
\includegraphics[width=1.0\textwidth]{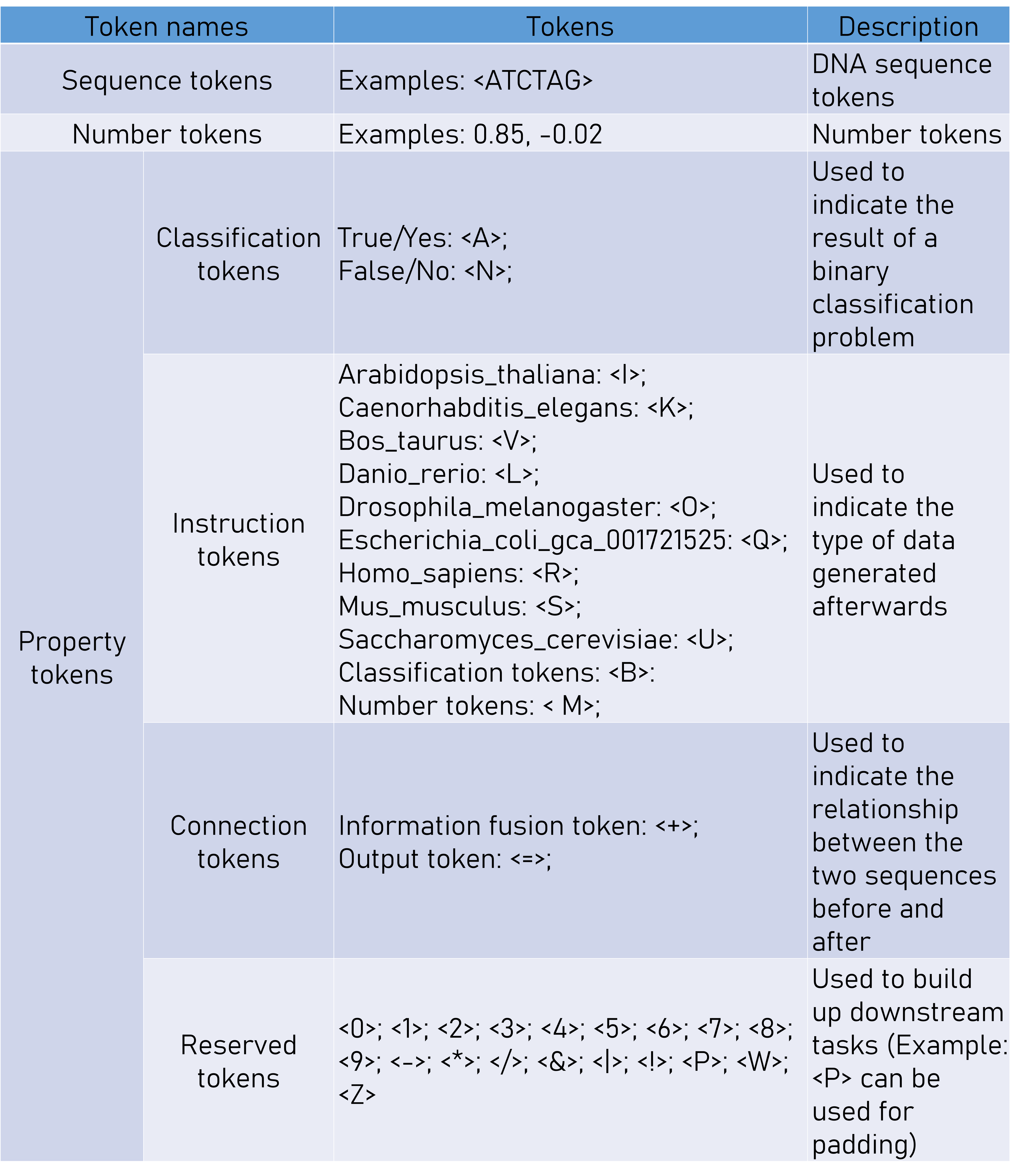}
\caption{All the tokens we used in our DNAGPT. We present the token name, the tokens (For cases with a large number of tokens, we only provided some examples.) and the token description.}
\label{s_token}
\end{figure*}

There are 6 categories of tokens in the token language of DNAGPT. The \textit{Sequence tokens} are the DNA sequences encoded with kmers tokenization strategy. For example, if we utilize 6-mers sampling and only consider the encoding of ’A, C, G, T, N’, then the total amount of discrete tokens are $5^6 + 5^5 +5^4 +5^3 +5^2 +5^1$ which is 19530. When comes to the Number tokens, we directly input the numbers into the \textit{Numerical embedding layer} and \textit{Regression head layer} to encode and decode them as the number tokens. For binary classification tasks, we utilize 'A' and 'N' to distinguish True from False. The \textit{Instruction tokens} are used to identify the input and output type of sequence. For DNA sequences from different species, we assign an instruction token for each species. Specifically, we also assign instruction tokens for Classification tasks and Numerical tokens which can prompt the model to generate corresponding types of tokens separately. In biological sequences, there is no natural logical relationship between tokens like in the natural language. In the design of DNAGPT tokens, to enable the model to understand the relationships among sequences, we design two connection tokens to guide the relationships between sequences before and after the connection tokens. Here, ’+’ represents the fusion of preceding and succeeding information, and ’=’ represents the cause-effect relationship, with the input being before ’=’ and the output being after ’=’. Finally, in order to better adapt to different types of downstream tasks, we also reserve some special tokens.

\subsection{Datasets}\label{dataset}

\subsubsection{Genomic signals and regions recognition}
 
The datasets used for genomic signals and regions recognition are cDNA data. We extracted both polyadenylation signals (PAS) and translation initiation sites (TIS) from four genomes. For the Homo sapiens (human) genome, the human assembly GRCh37 (also known as hg19) was employed, while the primary assembly GRCm38 was used for the Mus musculus (mouse) genome. The cDNA data for these genomes were sourced from the Mammalian Gene Collection (MGC). For the Bos taurus (bovine) genome, the assembly Bos\_taurus\_UMD\_3.1.1 was utilized, with the cDNA data being downloaded from the Ensembl organization. Finally, for the Drosophila melanogaster (fruit fly) genome, Release\_6 – annotation release Dmel\_Release\_6.01 was employed, and the cDNA data was obtained from FlyBase. The sampling method is as follows: first, locate the positions of GSRs, then extract 300 bps of sequence from both before and after the GSRs, and concatenate them together. It is important to note that the GSR motif will be removed during preprocessing to ensure that the model can recognize GSRs based solely on the information near the GSR motif, rather than the GSR itself. For the negative samples, the sampled sequences should satisfy the following requirements:

(1) Sequences with the same motifs but not related to polyadenylation and translation processes.

(2) Sequences are sampled from the chromosome whose average GC-content was nearest to the entire genome's average GC-content.

Consequently, negative data for human, mouse, bovine, and fruit fly were extracted from chromosomes 21, 13, 28, and X, respectively.

The amounts of positive samples for each dataset are shown in Table. \ref{s_data}.

\begin{table}[h]
\caption{Amounts of positive samples for different datasets.}
\begin{tabular}{@{}lccccc@{}}
\toprule
GSRs & Human & Mouse & Bovine & Fruit fly \\
\midrule
TIS & 28,244 & 25,205 & 17,558 & 30,283 \\

PAS(AATAAA) & 11,302 & 11,393 & 7,862 & 18,641 \\
ALL & 20,933 & 18,693 & 12,082 & 27,203 \\
\botrule
\end{tabular}
\label{s_data}
\end{table}

\subsubsection{Artificial human genomes generation}

For artificial human genomes generation, we utilized 1000 Genomes data \cite{10002015global} as the fine-tuning dataset. There are 2504 individuals (5008 haplotypes) in the dataset and the data we used is a dense 10000 SNP range/region from chromosome 15. When evaluating, the model produced 5000 sequences of SNPs for analysis. All our analyses were conducted on the generated data.

\subsubsection{mRNA expression levels prediction}

The dataset is composed of human protein-coding gene sequences located upstream and downstream of the transcription start site (TSS). The promoter of the gene is found in the sequence upstream of the TSS, while the exons and introns of the gene are found downstream. The input sequences are sourced from the Xpresso\cite{agarwal2020predicting}. In this dataset, the TSS positions were meticulously revised by the authors of Xpresso using Cap Analysis Gene Expression (CAGE) \cite{lizio2015gateways}, a technique for determining the actual TSS location. The Xpresso dataset consists of 18,377 promoters, divided into 16,377 for training, 1,000 for validation, and 1,000 for testing as mentioned in the Xpresso\cite{agarwal2020predicting}. The maximum length of a promoter's TSS sequence is set to 20,000 base pairs. The default sample range in xpresso is from 3000 to 13500 when DNAGPT can utilize the whole sequence. Additionally, the Xpresso DNA input includes half-life features that provide general information about the gene, such as gene length and the number of introns. The default feature input is an 8-bit array.

\subsection{Experiment details}

\subsubsection{Pre-training details}

\begin{figure}[!t]%
\centering
\includegraphics[width=0.95\textwidth]{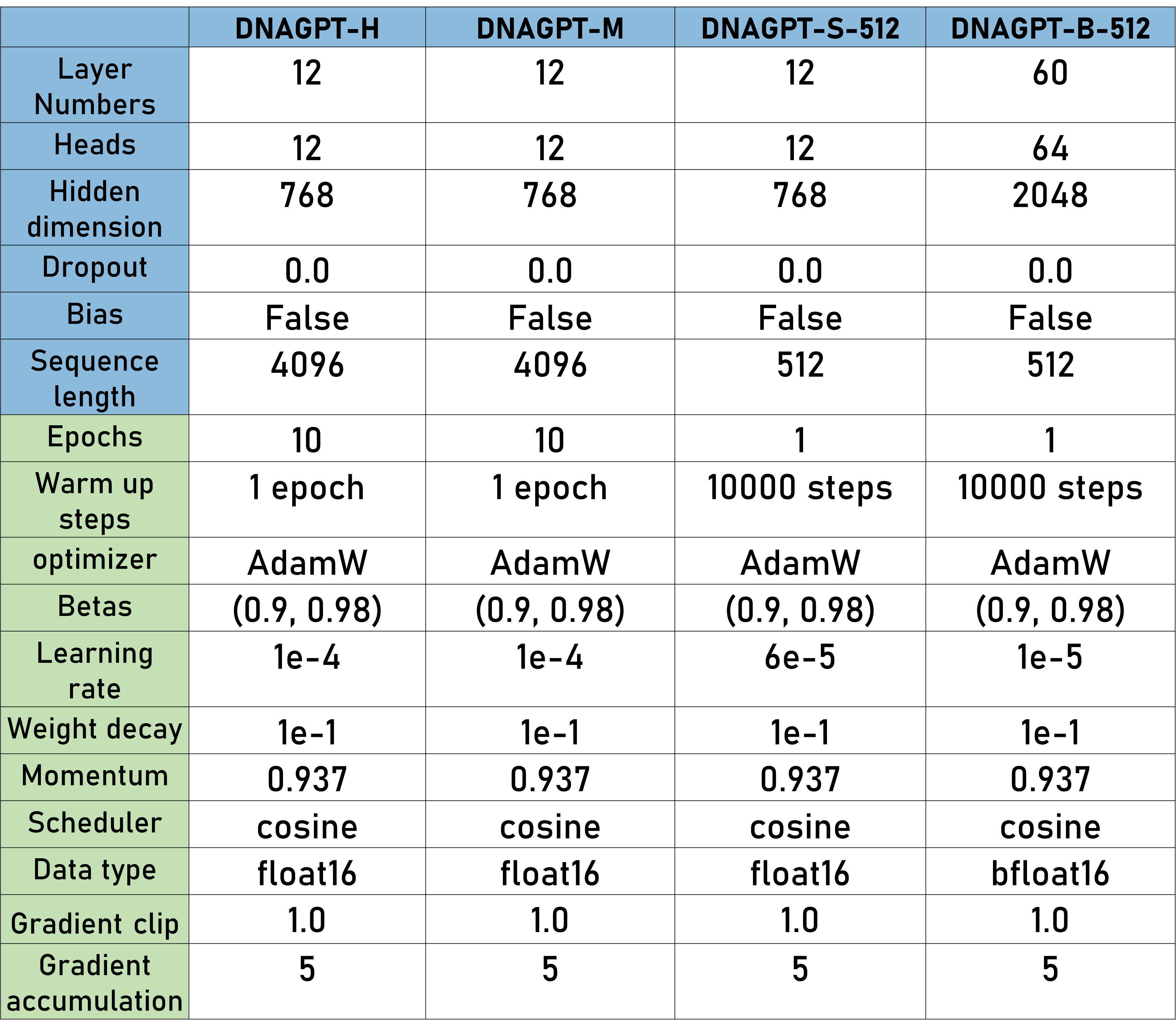}
\caption{Pre-training details of DNAGPTs.}\label{pretrain_detail}
\end{figure}

We show the detailed training information and hyper-parameters of our DNAGPTs in Figure \ref{pretrain_detail}. We utilize deepspeed \cite{rasley2020deepspeed} and FSDP in the pretraining process. We also utilize json data format to organize the data in order to accelerate the training of DNAGPT. DNAGPT-H, DNAGPT-M, DNAGPT-S-512 are pretrained on 8 $\times$ V100 GPUs and DNAGPT-B-512 is pretrained on 16 $\times$ V100 GPUs. 

\subsubsection{Fine-tune details}

\begin{figure}[!t]%
\centering
\includegraphics[width=0.95\textwidth]{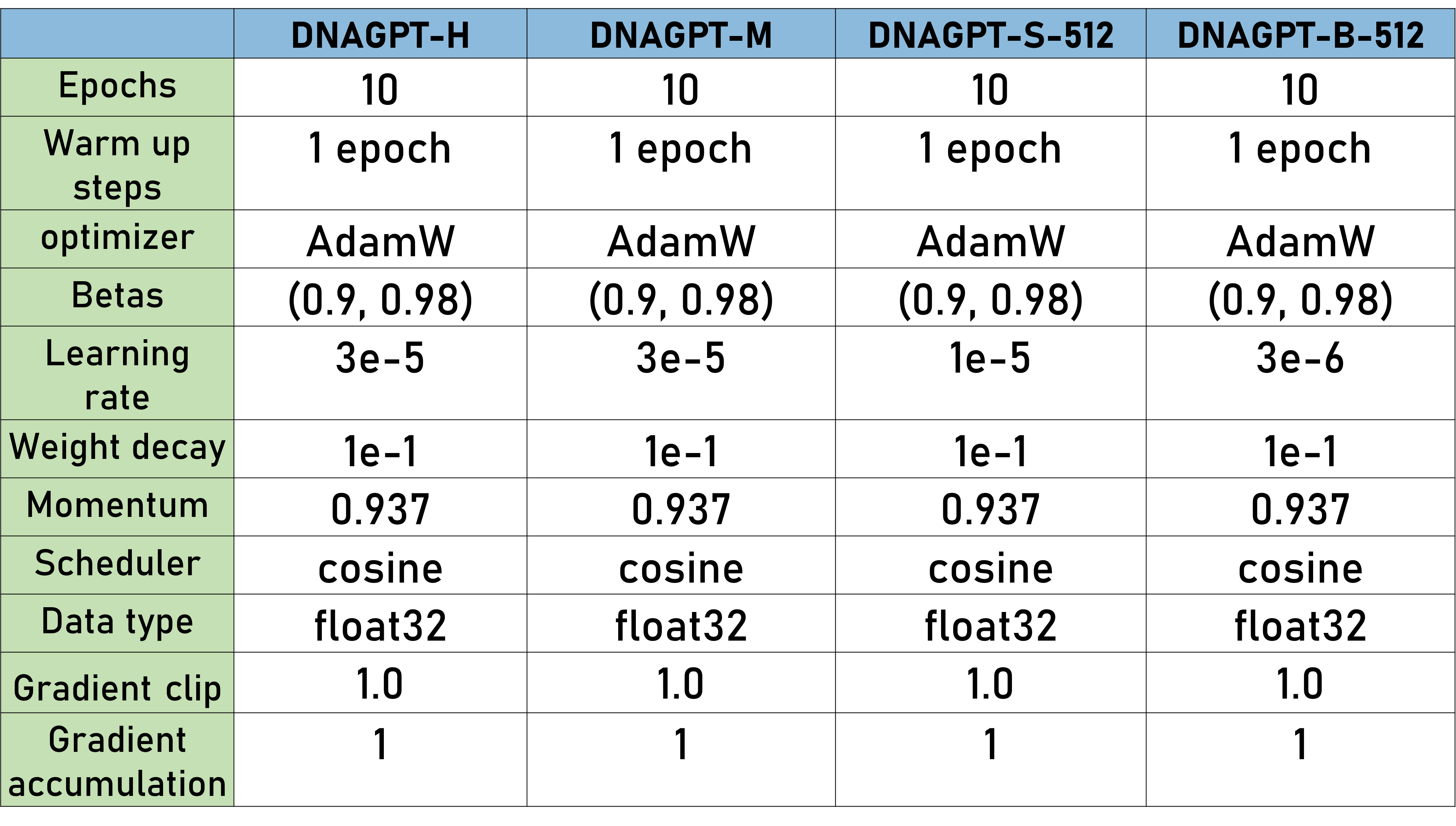}
\caption{Fine-tune details of DNAGPT on the GSR recognition task.}\label{finetune_detail}
\end{figure}

Specifically, we report the fine-tune details on GSR recognition dataset for each DNAGPT in Figure. \ref{finetune_detail}. We fine-tuned out model for 10 epochs on each dataset and decrease the learning rate to one-third of which in the pre-training time. In the mRNA prediction task and artificial human genomes generation, we use DNAGPT-H and DNAGPT-M and the settings remains the same as mentioned in Figure. \ref{finetune_detail}. Specifically, in the artificial human genomes generation task, we employed a post-processing stage to filter out sequences that do not meet the requirements by examining whether the predetermined stop signal token is in the correct position.

\end{document}